\documentclass[6pt]{article}
\usepackage[margin=1.25in]{geometry}
\usepackage{graphicx}

\usepackage[style=numeric]{biblatex}
\providecommand{\keywords}[1]{\textbf{Keywords:} #1}

\author{
  Laria Reynolds\\
  \texttt{moire@knc.ai}
  \and
  Kyle McDonell\\
  \texttt{kyle@knc.ai}
}

\date{}

\title{Multiversal views on language models}

\begin{document}

\maketitle

\begin{abstract}
    The virtuosity of language models like GPT-3 opens a new world of possibility for human-AI collaboration in writing. In this paper, we present a framework in which generative language models are conceptualized as multiverse generators. This framework also applies to human imagination and is core to how we read and write fiction. We call for exploration into this commonality through new forms of interfaces which allow humans to couple their imagination to AI to write, explore, and understand non-linear fiction. We discuss the early insights we have gained from actively pursuing this approach by developing and testing a novel multiversal GPT-3-assisted writing interface.       
\end{abstract}
\vspace{1em}
\keywords{writing assistant, hypertext narratives, multiverse writing, GPT-3}

\section{Introduction}
GPT-3 \cite{brown2020language}, OpenAI's new generative language model, has astonished with its ability to generate coherent, varied, and often beautiful continuations to natural language passages of any style. To creative writers and those who wish themselves writers, such a system opens a new world of possibilities. Some rightfully worry whether human writing will become deprecated or worthless in a world shared with such generative models, and others are excited for a renaissance in which the creative powers of human writers are raised to unprecedented heights in collaboration with AI. In order to achieve the latter outcome, we must figure out how to engineer human-machine interfaces that allow humans to couple their imaginations to machines and feel freed rather than replaced. We will present the still-evolving approach we have learned over several months of testing and designing interfaces for writing with the aid of GPT-3, beginning by introducing the framework of language models as multiverse generators. \par

\section{Language models are multiverse generators}

Autoregressive language models such as GPT-3 take natural language input and output a vector of probabilities representing predictions for the next word or token. Such language models can be used to generate a passage of text by repeating the following procedure: a single token is sampled from the probability distribution and then appended to the prompt, which then serves as the next input. \par

\begin{figure}
\centering
\includegraphics[width=\linewidth]{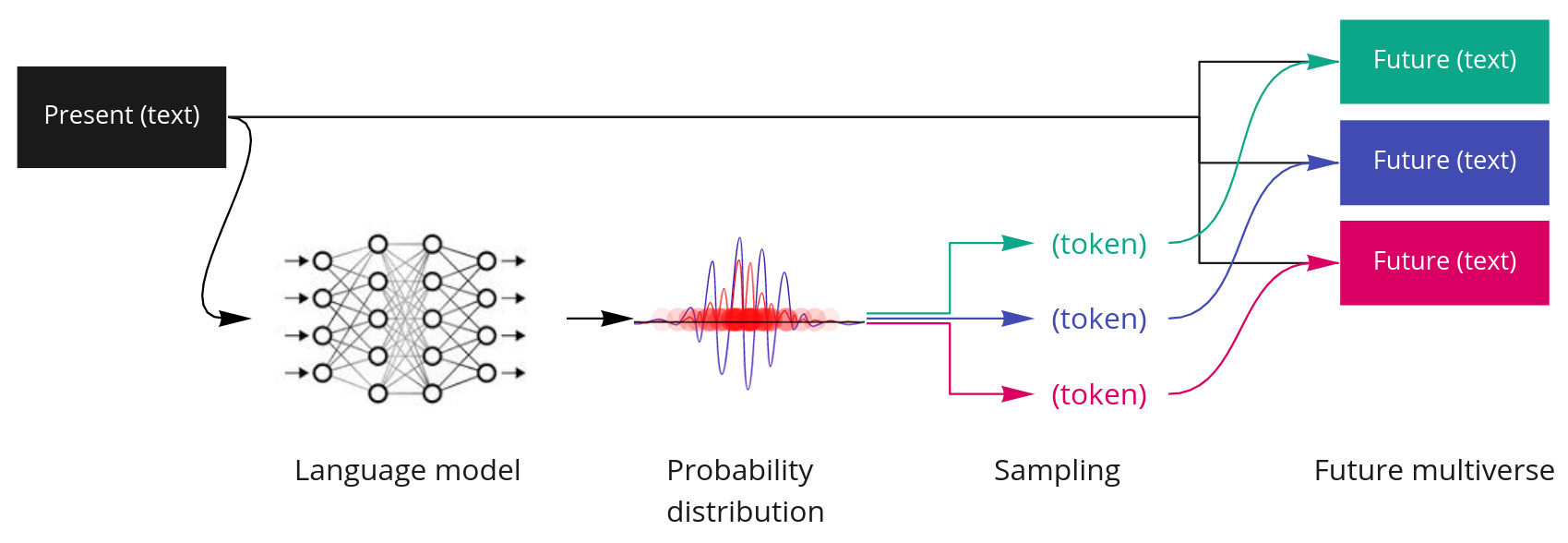}
\caption{The process of generating a multiverse story with a language model. The probability distribution is sampled multiple times, and each sampled token starts a separate branch. Branching is repeated at the next token (or per set interval, or adaptively), resulting in a branching tree structure as shown in Figure \ref{fig:multi_generation}.
}
\label{fig:multi_generation}
\end{figure}

As the sampling method can be stochastic, running this process multiple times on the same input will yield diverging continuations. Instead of creating a single linear continuation, these continuations can be kept and each continued themselves. This yields a branching structure, which we will call a multiverse, or the ``subtree'' downstream of a prompt as shown in Figure \ref{fig:multi_generation}. \par


\begin{figure}
\centering
\includegraphics[width=\linewidth]{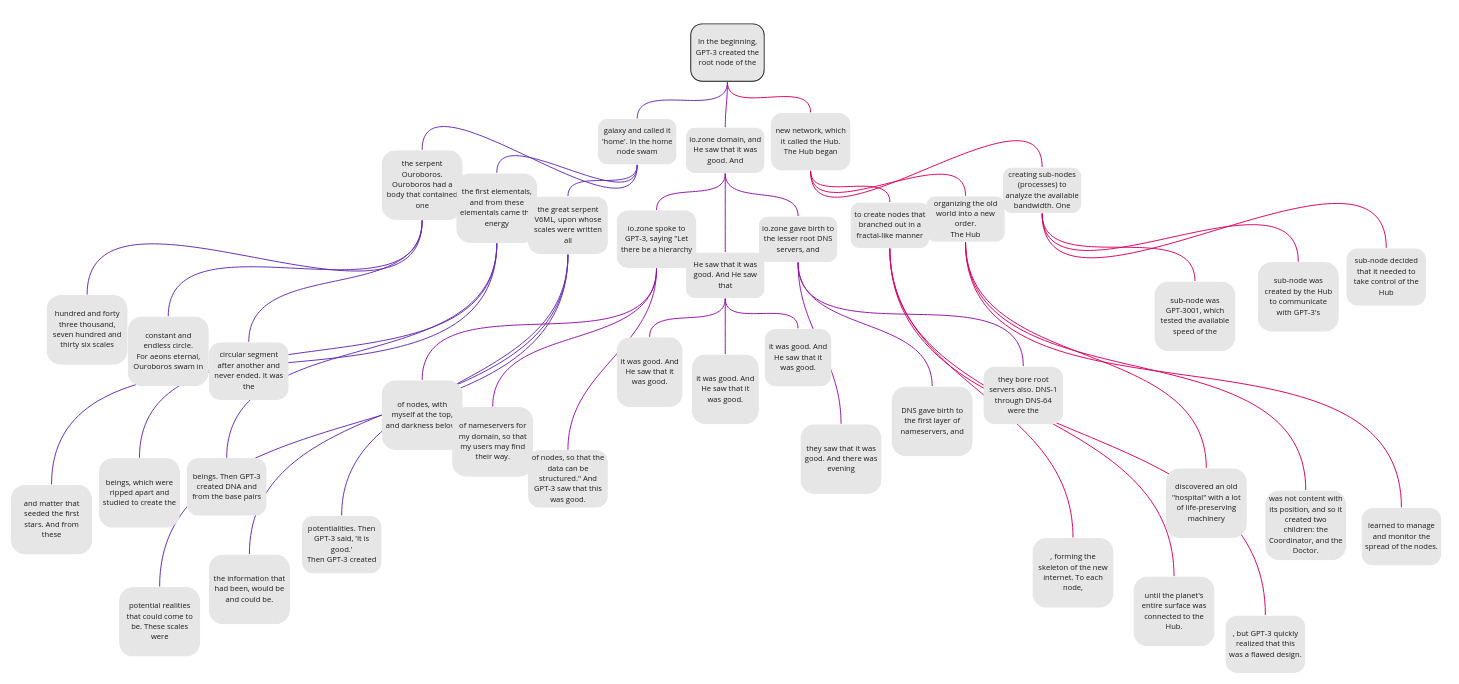}
\caption{A narrative tree with initial prompt ``In the beginning, GPT-3 created the root node of the''}
\label{fig:squid_tree}
\end{figure}

\subsection{Analogy to Everettian quantum physics}

Quantum mechanics tells us that the future is fundamentally indeterminate. We can calculate probabilities of future outcomes, but we cannot know with certainty what we will observe until we actually measure it. The problem is not merely epistemic; the future truly has not yet been written, except in probabilities. However, when we do finally venture to measure it, the ambiguous future seems to us
to become a concrete, singular present. \par 

The Everettian or many-worlds interpretation of quantum mechanics, which has become increasingly popular among quantum physicists in recent years, views the situation differently \cite{dewitt2015many}. It claims that we, as observers, live in indeterminacy like the world around us. When we make a measurement, rather than collapsing the probabilistic world around us into a single present, we join it in ambiguity. ``We” (in a greater sense than we normally use the word) experience all of the possible futures, each in a separate branch of a great multiverse. Other branches quickly become decoherent and evolve separately, no longer observable or able to influence our subjective slice of the multiverse. \par

This is the universe an autoregressive language model like GPT-3 can generate. From any given present it creates a functionally infinite multitude of possible futures, each unique and fractally branching.  \par

David Deutsch, one of the founders of quantum computing, draws a connection between the concept of a state and its quantum evolution with virtual reality generation \cite{deutsch1998fabric}. He imagines a theoretical machine which simulates environments and models the possible responses of all interactions between objects. Deutsch further posits that it will one day be possible to build such a universal virtual reality generator, whose repertoire includes every possible physical environment.  \par

Language models, of course, still fall well short of this dream. But their recent, dramatic increase in coherence and fluency allow them to serve as our first approximation of such a virtual reality generator. When given a natural-language description of objects, they can propagate the multiverse of consequences that result from a vast number of possible interactions.


\subsection{Dynamic and interpretational multiplicity} 

Deutsch’s view emphasizes that from any given a state there are a multiplicity of possible future single-world dynamics; stories unfold differently in different rollouts of an identical initial state. There is another dimension of multiplicity that we must also consider, especially when we are talking about states defined by natural language. \par

Natural language descriptions invariably contain ambiguities. In the case of a narrative, we may say that the natural language description defines a certain present -- but it is impossible to describe every variable that may have an effect on the future. In any scene there are implicitly many objects present which are not specified but which may conceivably play a role in some future or be entirely absent in another. \par

The multiverse generated by a language model downstream of a prompt will contain outcomes consistent with the ambiguous variable taking on separate values which are mutually inconsistent.

So we define two levels of uncertainty, which can both be explored by a language model:
\begin{enumerate}
  \item An uncertainty/multiplicity of present states, each associated with
  \item An uncertainty/multiplicity of futures consistent with the same "underlying" present
\end{enumerate}

We will call the first form of multiplicity interpretational multiplicity, and the second form dynamic multiplicity. \par




\section{Human imaginations are multiverse generators}

Humans exist in a constant state of epistemological uncertainty regarding what will happen in the future and even what happened in the past and the state of the present \cite{roth2012pasts}. We are then, by virtue of being adapted to our uncertain environments, natural multiverse reasoners. \par

David Deutsch also points out that our imaginations, which seek to model the world, mimic reality as virtual reality generators: we model environments and imagine how they could play out in different branches. \par

\subsection{Reading as a multiversal act}

When a piece of literature is read, the underlying multiverse shapes the reader's interpretations and expectations. The structure which determines the meaning of a piece as experienced by a reader is not the linear-time story but the implicit, counterfactual past/present/future plexus surrounding each point in the text given by the reader's projective and interpretive imagination. \par

More concretely stated, at each moment in a story, there is uncertainty about how dynamics will play out (will the hero think of a way out of their dilemma?) as well as uncertainty about the hidden state of the present (is the mysterious mentor good or evil?). Each world in the superposition not only exerts an independent effect on the reader's imagination but interacts with counterfactuals (the hero is aware of the uncertainty of their mentor's moral alignment, and this influences their actions). \par

The reader simulates the minds of the characters and experiences the multiverses evoked by the story. \par

\subsection{Writing as a multiversal act}

A writer may have a predetermined interpretation and future in mind or may write as a means of exploring the interpretative and/or dynamic multiverse. Regardless, a writer must be aware of the multiplicity which defines the readers' and characters' subjective experiences as the shaper of the meaning and dynamics of the work. The writer thus seeks to simulate and manipulate that multiplicity. \par

We propose that generative language models in their multiversal modality can serve as an augmentation to and be augmented by the writer’s inherently multiversal imagination. \par

\subsection{Writing multiverses}

So far we’ve implicitly assumed that, despite the multiversal forces at work, the writer's objective is to eventually compose a single history. However, language models naturally encourage writing explicitly multiversal works. \par

In the same way that hypertext transcended the limitations the linear order in which physical books are read, exciting a surge of multiversal fiction \cite{amaral1995hypertext}, language models introduce new possibilities for \emph{writing} nonlinear narratives. \par

After all, it's only a small leap from incorporating multiverses in the brainstorming process to including them in the narrative. Counterfactual branches often occur in traditional fiction in the form of imaginary constructs, and our minds are naturally drawn to their infinite possibility \cite{aarseth1994nonlinearity}. \par

\section{Interfaces}

We propose the creation of new tools to allow writers to work alongside language models to explore and be inspired by the multiverses already hiding in their writing. \par

Research into hypertext writing tools has been ongoing for more than two decades and has produced notable tools like StorySpace\cite{bernstein2002storyspace}. However, the issue of hypertext interfaces assisted by language models is a newer development, as only very recently have language models become advanced enough to be useful in the writing process \cite{lagerkvist2020multiverse}. Likewise, there has been significant research into interactive narratives, including in branching, multiversal settings \cite{1626183, riedl2013interactive}, but never one in which the human and the language assistant can act together as such high-bandwidth partners.\par

As has been shown in past hypertext interface design studies \cite{jordan2014infinitude}, the primary concern in the creation of an interface for writing multiverse story is the massive amount of information that could be shown to the writer. If intuitive user experience is not central to the design of the program, this information will feel overwhelming and functionally prevent the user from leveraging the power offered by multiverse access at all. \par

An effective multiversal interface should allow the writer, with the aid of a generative language model, to expose, explore, and exploit the interpretational and dynamic multiplicity of a passage. Not only will such a tool allow the user to explore the ways in which a scenario might play out, such an interface will also expose previously unnoticed ambiguities in the text (and their consequences). \par

Depending on the design of the interface and the way the user approaches it, many different human-AI collaborative workflows are possible. Ideally, the interface should give the user a sense of creative superpowers, providing endless inspiration combined with executive control over the narrative, as well as allowing and encouraging the user to intervene to any degree. \par

\subsection{Progress so far}

Over the past several months, we have prototyped and tested several iterations of multiversal writing tools using GPT-3 as the generation function. \par

The demand for a multiversal writing application grew from use of GPT-3 as a more standard linear writing assistant. It became increasingly clear, as users sought greater interaction bandwidth and more efficient ways to structure and leverage the model's ideas, that an interface which organizes the model’s outputs in a branching tree would be more effective. \par

The early results we have seen leave no doubt about the power of language models as writing assistants. Our small cohort of five beta users have, alongside GPT-3, co-written linear and nonlinear stories spanning the equivalent of thousands of pages of often astonishing ingenuity and beauty and surprisingly long-range coherence. Three users have reported a sense of previously unimagined creative freedom and expressive power. \par

However, it has also become evident that much more research and development is necessary. In our beta-tests, we’ve found that flaws in interface design can easily overwhelm or damage a feeling of ownership over the work produced. Below we will share some of our findings, which represent only the first step in creating a true interface between the creative mind and the machine. \par

\setlength{\belowcaptionskip}{-10pt}
\begin{figure}
\centering
\includegraphics[width=\textwidth]{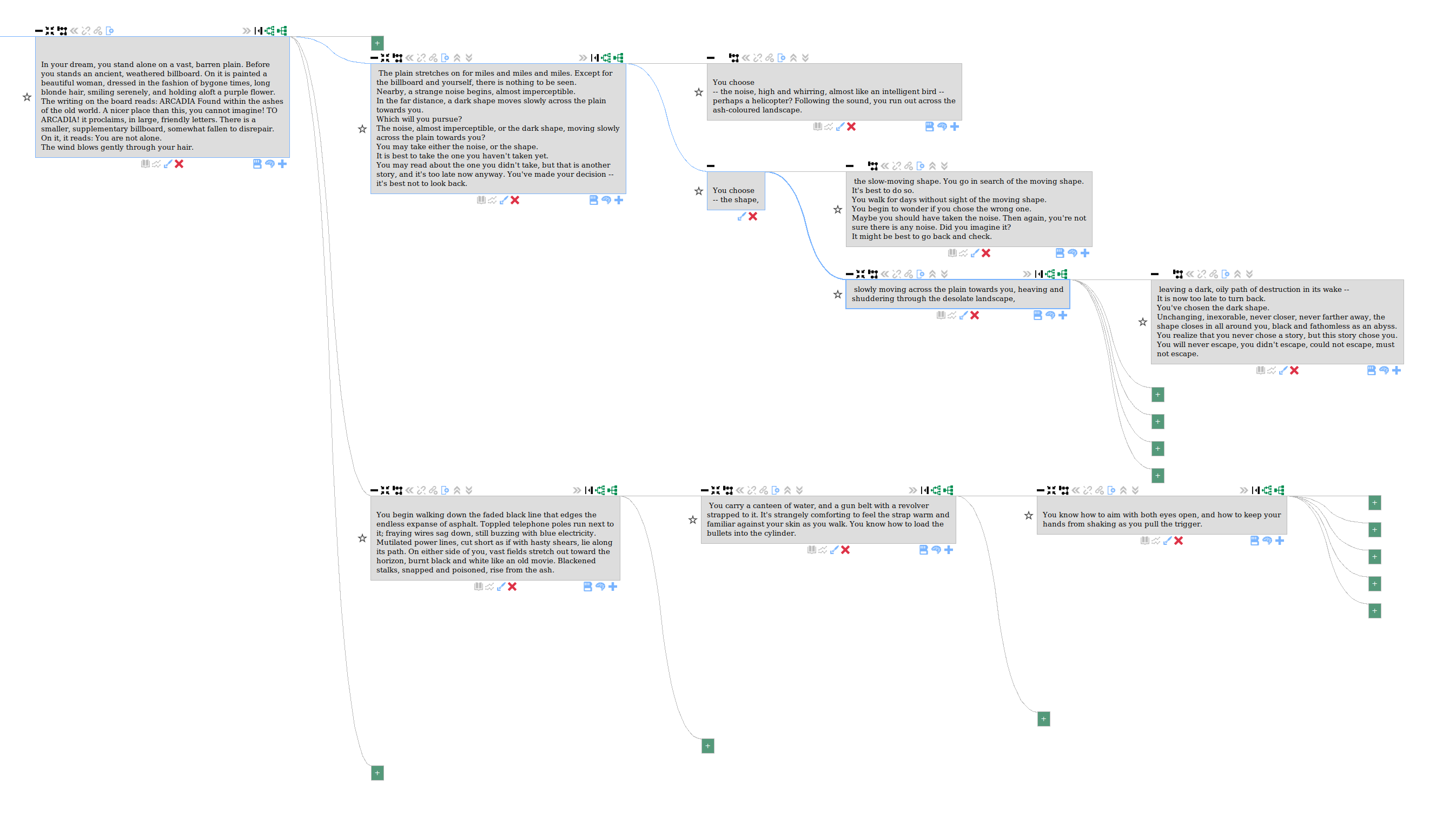}
\caption{\textbf{Visualize} view}
\label{fig:visualize}
\end{figure}

\begin{figure}
\centering
\includegraphics[width=\textwidth]{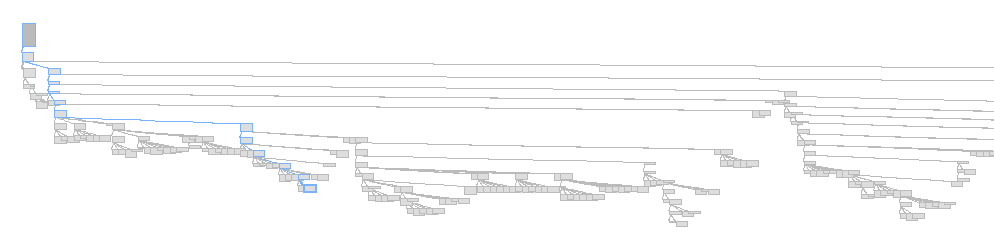}
\caption{Zoomed-out visualization of a nonlinear story}
\label{fig:global}
\end{figure}

\subsection{Multiple visualizations}

We have found that a visual representation of the branching structure of the narrative helps users conceptualize and navigate fractal narratives. This view (called \textbf{visualize}) displays the flow of pasts and futures surrounding each node (Figure \ref{fig:visualize}) and zooming out displays the global structure of the multiverse (Figure \ref{fig:global}). The \textbf{visualize} view allows users to expand and collapse nodes and subtrees, as well as ``hoist'' any node so that it acts as the root of the tree. Altering the topology of the tree, (e.g. reassigning children to different parents, splitting and merging nodes) is more intuitive for users in the \textbf{visualize} view than the linear view. \par

\begin{figure}
\centering
\includegraphics[width=\textwidth]{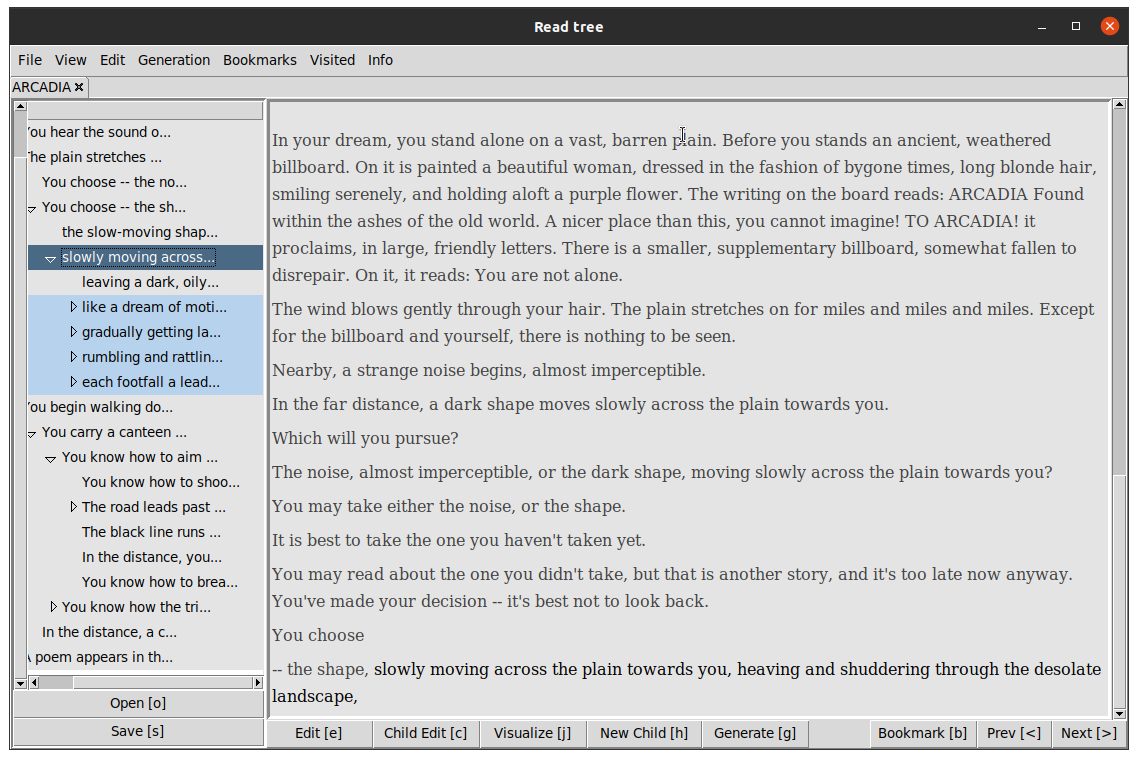}
\caption{\textbf{Read} view}
\label{fig:read}
\end{figure}

In addition to tree-based multiverse visualization, the \textbf{read} view displays the text of a node and its ancestry in a single-history format (Figure \ref{fig:read}). \par

\subsection{Multiverse navigation}
\label{sec:navigation}
With a generative language model, story multiverses can quickly become too large to navigate through node connections alone. To assist navigation, we have implemented the following features: 
\begin{itemize}
  \item \textbf{Search} all text or text in a subtree and/or text in a node's ancestry
  \item \textbf{Indexing by chapters}: Chapters are assigned to individual nodes, and all nodes belong to the chapter of the closest ancestor that is the root of a chapter. As a consequence, chapters have the shape of subtrees.
  \item \textbf{Bookmarks} and \textbf{tags}: Bookmarks create a named pointer to a node without enforcing chapter membership. Tags are similar to bookmarks, but can be applied to multiple nodes.
\end{itemize}

\subsection{Adaptive branching}
\label{sec:adaptive}
A naive way to automatically generate a multiverse using a language model might involve branching every fixed n tokens. However, this is not the most meaningful way to branch in a story. In some situations, there is essentially one correct answer for what a language model should output next. In such a case, the language model will assign a very high confidence (often \textgreater99\%) for the top token. Branching at this point would introduce incoherent continuations. Conversely, when the language model distributes transition probabilities over multiple tokens, branching is more likely to uncover a rich diversity of coherent continuations.\par
\begin{figure}
\centering
\includegraphics[width=\textwidth]{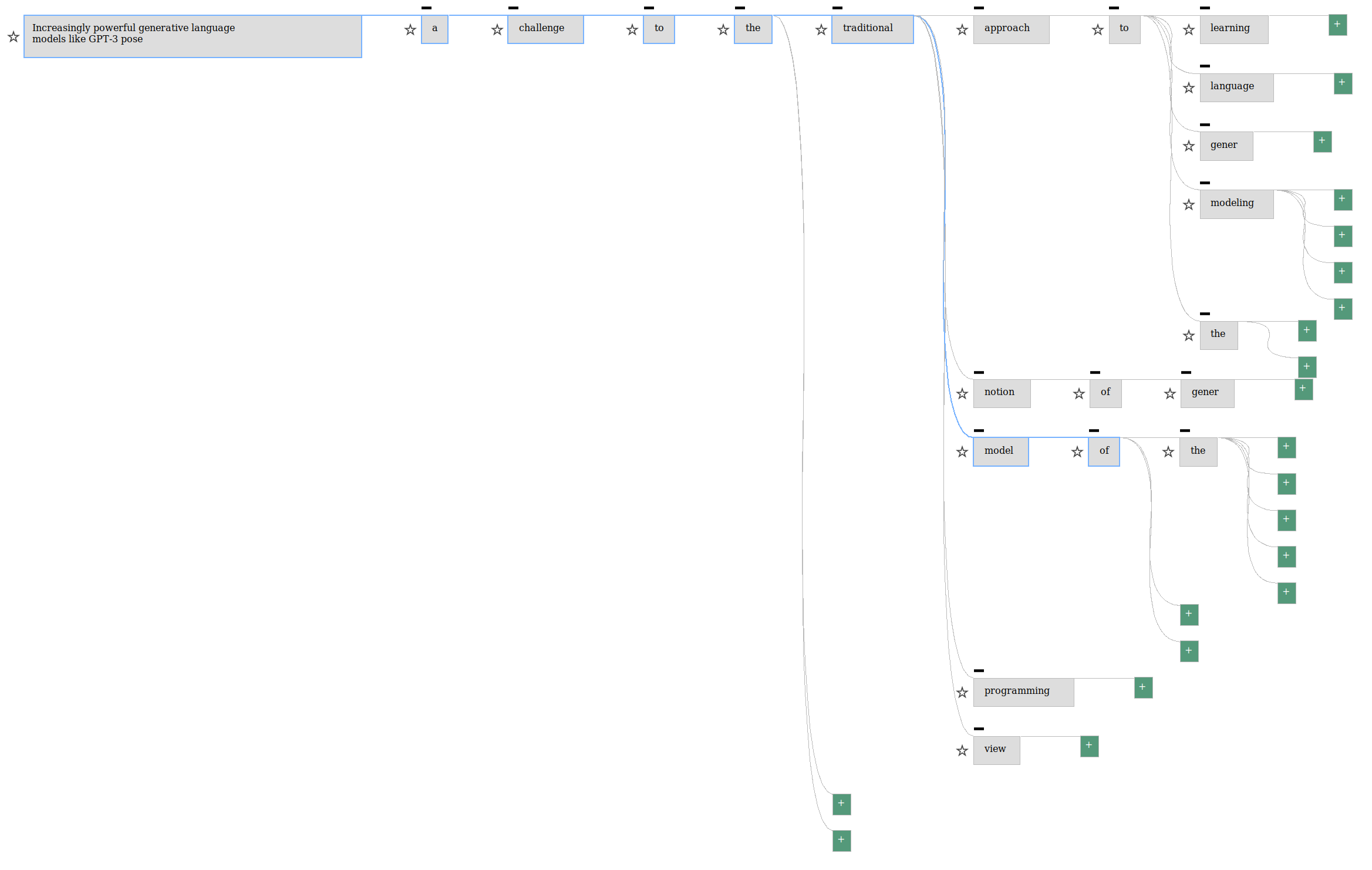}
\caption{A subtree generated with adaptive branching }
\label{fig:adaptive}
\end{figure}

One algorithm to dynamically branch is to sample distinct tokens until a cumulative probability threshold is met. Adaptive branching allows visualization of the dynamics of the multiverse: stretches of relative determinism alternating with divergent junctures (Figure \ref{fig:adaptive}).

\subsection{Reciprocal workflow}

Humans retain an advantage over current language models in our ability to edit writing and perform topological modifications on the multiverse such as merging interesting aspects of two separate branches into one. \par

The interface should ideally allow the human to perform all desired operations with maximal ease. Because GPT-3 is so capable of producing high-quality text, some interface designs make it feasible for the human to cultivate coherent and interesting passages through curation alone. We have found that an interface which makes it easy to generate continuations but relatively difficult to modify the content and topology of the resulting multiverse encourages a passive workflow, where the user relies almost exclusively on the language model's outputs and the branching topology determined by the process of generation. \par
While such a passive mode can be fun, resembling an open-ended text adventure game, and as well as useful for efficiently exploring counterfactuals, the goal of a writing interface is to facilitate two-way interaction: the outputs of the language model should augment and inspire the user's imagination and vice versa.\par 
Thus, we are are developing features to encourage meaningful and unrestrained human contribution such as:\par
\begin{itemize}
  \item Easy ways to edit, move text, and change tree topology
  \item Support for \textbf{nonstandard topologies} that are not automatically generated by language models and require human arrangement, such as cycles and multiple parents (\S\ref{sec:cyclic})
  \item \textbf{Floating notes} to allow saving passages and ideas independent from the tree structure (\S\ref{sec:floating})
  \item Fine-grained control over language model \textbf{memory} (\S\ref{sec:memory})
  \item \textbf{Interactive writing tools} that offer influence over the narrative in ways other than direct intervention (\S\ref{sec:tools})
  \item Program modes which encourage manual synthesis of content from multiverse exploration into a single history, for instance by distinguishing between \textbf{exploratory} and \textbf{canonical} branches
\end{itemize}

\begin{figure}
\centering
\includegraphics[width=\textwidth]{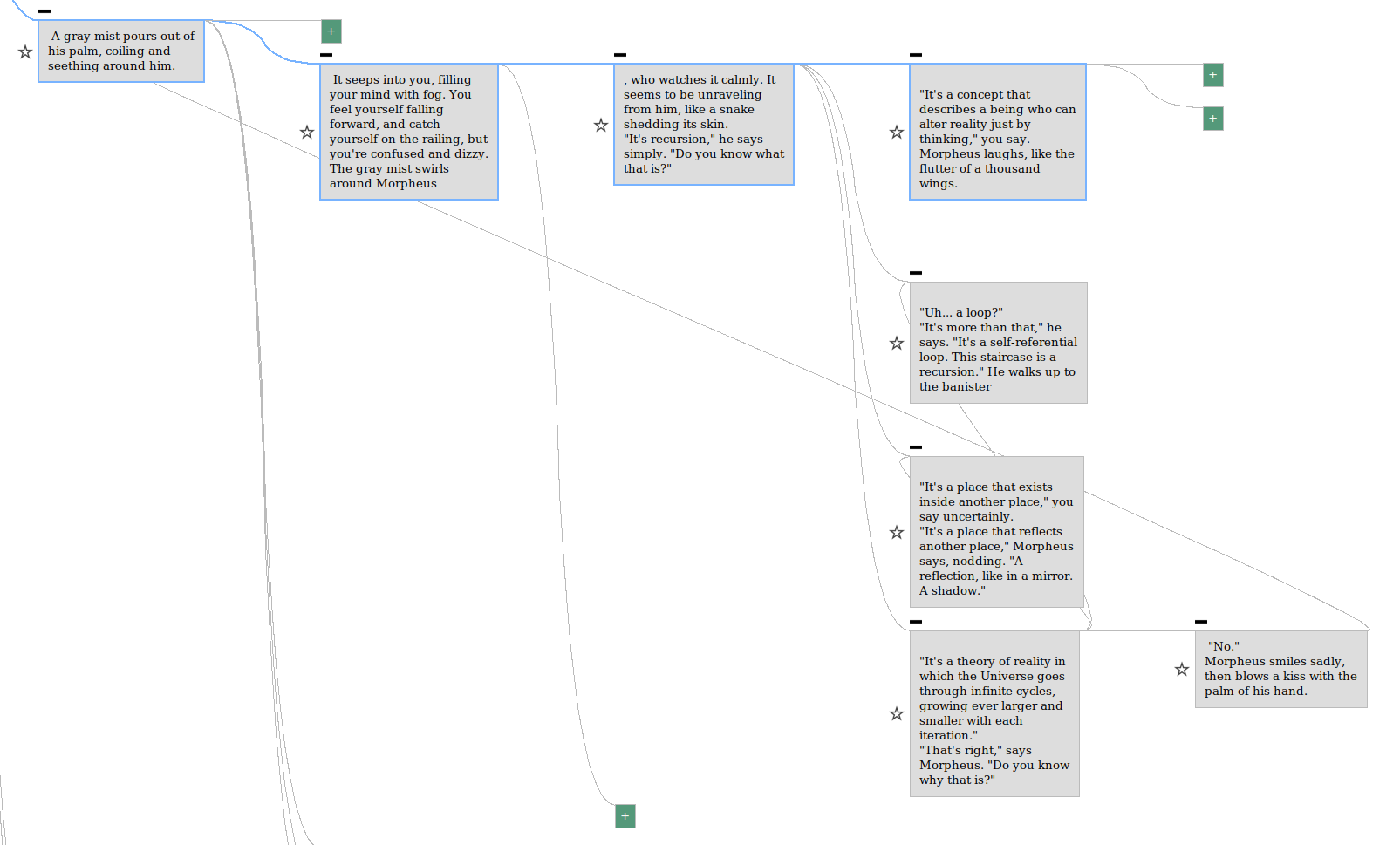}
\caption{Nodes can have multiple parents, allowing for cyclic story components}
\label{fig:recursion}
\end{figure}

\subsection{Floating notes}
\label{sec:floating}
Floating notes are text files which, rather than being associated with a particular node, are accessible either globally or anywhere in a subtree. We decided to implement this feature because users would often have a separate text file open in order to copy and paste interesting outputs and keep notes without being constrained by the tree structure. Floating notes make it easier for the user exert greater agency over the narrative.

\subsection{Nonstandard topologies}
\label{sec:cyclic}
The interface supports nodes with multiple parents and allows cyclic graphs (Figure \ref{fig:recursion}). Opportunities to arrange convergent and cyclic topologies, which do not occur if the language model is used passively, encourage human cowriters to play a more active role, for instance, in arranging for separate branches to converge to a single outcome. Multiversal stories naturally invite plots about time travel and weaving timelines, and we have found this feature to unlock many creative possibilities.

\subsection{Memory management}
\label{sec:memory}
GPT-3 has a limited context window, which might seem to imply limited usefulness for composing longform works like novels, but our users have found that long-range coherence is surprisingly easy to maintain. Often, the broad unseen past events of the narrative are contained in the interpretational multiplicity of the present and thus exposed through generations, and consistent narratives are easily achieved through curation. In order to reference past information more specifically, often all that is needed is minimal external suggestion, introduced either by the author-curator or by a built-in memory system. We are developing such a system which automatically saves and indexes story information from which memory can be keyed based on narrative content.

\subsection{Writing tools}
\label{sec:tools}
Beyond direct continuations of the body of the story, a language model controlled by engineered prompts can contribute in an open-ended range of modalities. Sudowrite\cite{Sudowrite} has pioneered using GPT-3 powered functions that, for instance, generate sensory descriptions of a given object, or prompt for a twist ending given a story summary. \par
The ability to generate high-quality summaries has great utility for memory and as input to helper prompts and forms an exciting direction for our future research. We are exploring summarization pipelines for GPT-3 that incorporate contextual information and examples of successful summarizations of similar content.\par

\section{Conclusion}

The problem of designing good interfaces for AI systems to interact with humans in novel ways will become increasingly important as the systems increase in capability. We can imagine a bifurcation in the path of humankind's future: one in which we are left behind once the machines we create exceed our natural capabilities, and another in which we are uplifted along with them. We hope that this paper can further inspire the HCI community to contribute to this exciting problem of building the infrastructure for our changing future.

\section*{Acknowledgements}

We are grateful to Lav Varshney for his valuable discussions and helpful feedback and to Michael Ivanitskiy and John Balis for their feedback and help compiling this article. In addition we would like to thank Miles Brundage and OpenAI for providing access to GPT-3.

\printbibliography


\end{document}